\documentclass[journal]{IEEEtran}
\usepackage[numbers,sort&compress]{natbib}
\usepackage{url}
\usepackage[cmex10]{amsmath}
\usepackage{graphicx,natbib,amssymb,lineno,color}
\usepackage{subfigure}
\usepackage{slashbox}
\usepackage{color}
\usepackage{xcolor}
\usepackage{listings}
\usepackage{booktabs}
\usepackage{threeparttable}
\usepackage{tabularx}
\usepackage{multicol}
\usepackage{multirow}

\newtheorem{remark}{Remark}

\usepackage{amsmath}
\usepackage{enumerate}
\usepackage{breqn}
\usepackage{setspace}

\usepackage[justification=centering]{caption}
\usepackage{algorithm}
\usepackage{algpseudocode}
\usepackage{setspace}
\renewcommand{\baselinestretch}{1.03}


\ifCLASSINFOpdf
\else
\fi

\begin{document}
\title{Machine Learning on Cloud with Blockchain: A Secure, Verifiable and Fair Approach to Outsource the Linear Regression}
\author{Hanlin~Zhang, Peng~Gao, Jia~Yu, Jie~Lin, and Neal N.~Xiong, \emph{Senior Member, IEEE}
	\thanks{
		Hanlin~Zhang, Peng~Gao and Jia~Yu are with the College of Computer Science and Technology, Qingdao University, Qingdao 266071, China, also with the State Key Laboratory of Cryptology, Beijing 100878, China (Emails: hanlin@qdu.edu.cn; 1033986392@qq.com; qduyujia@gmail.com).

        Hanlin~Zhang is also with the business school, Qingdao University, Qingdao 266071, China.

		Jie~Lin is with the School of Electronic and Information Engineering, Xi’an Jiaotong University, Xi’an 710049, China (E-mail: jielin@mail.xjtu.edu.cn).
		
		Neal~N.~Xiong is with the Department of Mathematics and Computer Science, Northeastern State University, USA. (E-mail:xiongnaixue@gmail.com).
 		
In this paper, J. Yu is the corresponding author.
 }}

\maketitle

\begin{abstract}
Linear Regression (LR) is a classical machine learning algorithm which has many applications in the cyber physical social systems (CPSS) to shape and simplify the way we live, work and communicate. This paper focuses on the data analysis for CPSS when the Linear Regression is applied. The training process of LR is time-consuming since it involves complex matrix operations, especially when it gets a large scale training dataset In the CPSS. Thus, how to enable devices to efficiently perform the training process of the Linear Regression is of significant importance. To address this issue, in this paper, we present a secure, verifiable and fair approach to outsource LR to an untrustworthy cloud-server. In the proposed scheme, computation inputs/outputs are obscured so that the privacy of sensitive information is protected against cloud-server. Meanwhile, computation result from cloud-server is verifiable. Also, fairness is guaranteed by the blockchain, which ensures that the cloud gets paid only if he correctly performed the outsourced workload. Based on the presented approach, we exploited the fair, secure outsourcing system on the Ethereum blockchain. We analysed our presented scheme on theoretical and experimental, all of which indicate that the presented scheme is valid, secure and efficient.

\end{abstract}


\IEEEpeerreviewmaketitle

\section{Introduction}
Cyber physical social systems (CPSS), as an emerging paradigm, is providing efficient, convenient, and personalized services and benefiting human lives. Machine Learning (ML) techniques have been widely implemented in the CPSS for providing new predictive models for large scale data analysis in various applications. Linear Regression (LR) is a classical supervised learning algorithm, which is widely used to establish the relationship between the target variable and the input variable based on a trained model. The training process of the linear regression involves the matrix multiplications and the matrix inversion, which are time-consuming operations. On the other hand, in the age of big data, the LR is often applied to a large-scale training dataset, which will require unaffordable computation power for individuals. Thus, how to enable devices to efficiently perform the training process of the linear regression is a critical problem.

In the age of cloud computing, outsourcing training process to an untrusted cloud can be an alternative solution. Cloud computing is providing fast and secure computing and data storage services over the internet. With cloud computing, users can avoid the upfront cost and complexity of owning and maintaining their own hardware, and instead simply pay for what they use. Although outsourcing the heavy workload to a cloud server has many advantages, it also brings some new challenges~\cite{ZYZ19,YRW16}. First, the outsourced data might include private information such as patients' health records, which should not be leaked to a public cloud. How to enable the cloud to perform the computation on privacy-preserved input is a critical challenge. Second, cloud server may return invalid computation outputs prepensely or unconsciously. The incorrect results might be caused by a software bug, a malicious attack on the cloud, or financial incentives to save computation power. Thus, how to enable outsourcer to detect malicious behavior from cloud servers is another challenge. Third, if cloud server performs computation task before outsourcer paying the services fee, the outsourcer might not pay after receiving the results. If the outsourcer pays the service fee first, the cloud might not conduct the computation and return random and invalid results. Thus, how to guarantee fairness for both the cloud server and the outsourcer is the last challenge. As far as we know, no existing research achieves secure, verifiable, and fair outsourcing for the linear regression.

To address the above challenges, in this article, we make research on how to safely implement a linear regression model on an untrustworthy cloud, in a way that guarantees fairness for both parties. To be specific, we propose a new elementary transformation based technique to obscure the computation input and output. To achieve fairness, taking advantage of the decentralized, traceable nature of the blockchain, we employ the blockchain as the middleman, which verifies the calculation results and guarantees the fairness. Notice that the verification process of secure outsourcing schemes often involves private data of the outsourcer, while the blockchain is a public ledger that anyone can have access to the data on it. Thus, we propose a verification method that does not involve any private information (i.e., the proposed outsourcing scheme is publicly verifiable). Based on the designed scheme, we implement the fair, secure outsourcing system on the Ethereum blockchain~\cite{wood2014ethereum}, which includes a verification system and a payment system. We develop the smart contract as well as the graphical user interface and introduce the implementation details. To evaluate our presented scheme, we analyze the correctness, security and efficiency of our approach on theoretical. Also, we perform experiment to assess the efficiency of our presented scheme.

The rest of the paper is organized as follows: Section~\ref{sec:two} introduces some essential preliminaries for the proposed scheme. Section~\ref{sec:three} provides the system model. Section~\ref{sec:four} describes the design rationale, the generic framework and the detailed scheme. Also, we analyses the correctness, security and efficiency of the presented scheme. Section~\ref{sec:six} introduces the implementation of the developed system in detail. Section~\ref{sec:seven} evaluates the practical performance of the proposed scheme through experiments. In Section~\ref{sec:app}, we discuss some possible applications where our proposed algorithm can be applied. Section~\ref{sec:eight} overviews the related work. Finally, Section~\ref{sec:nine} draws conclusions to the paper.

\section{PRELIMINARIES}	\label{sec:two}
In this section, we introduce some background knowledge of Linear Regression, Blockchain and Ethereum smart contract.
\begin{figure*}
	\centering
	\includegraphics[width=0.85\linewidth]{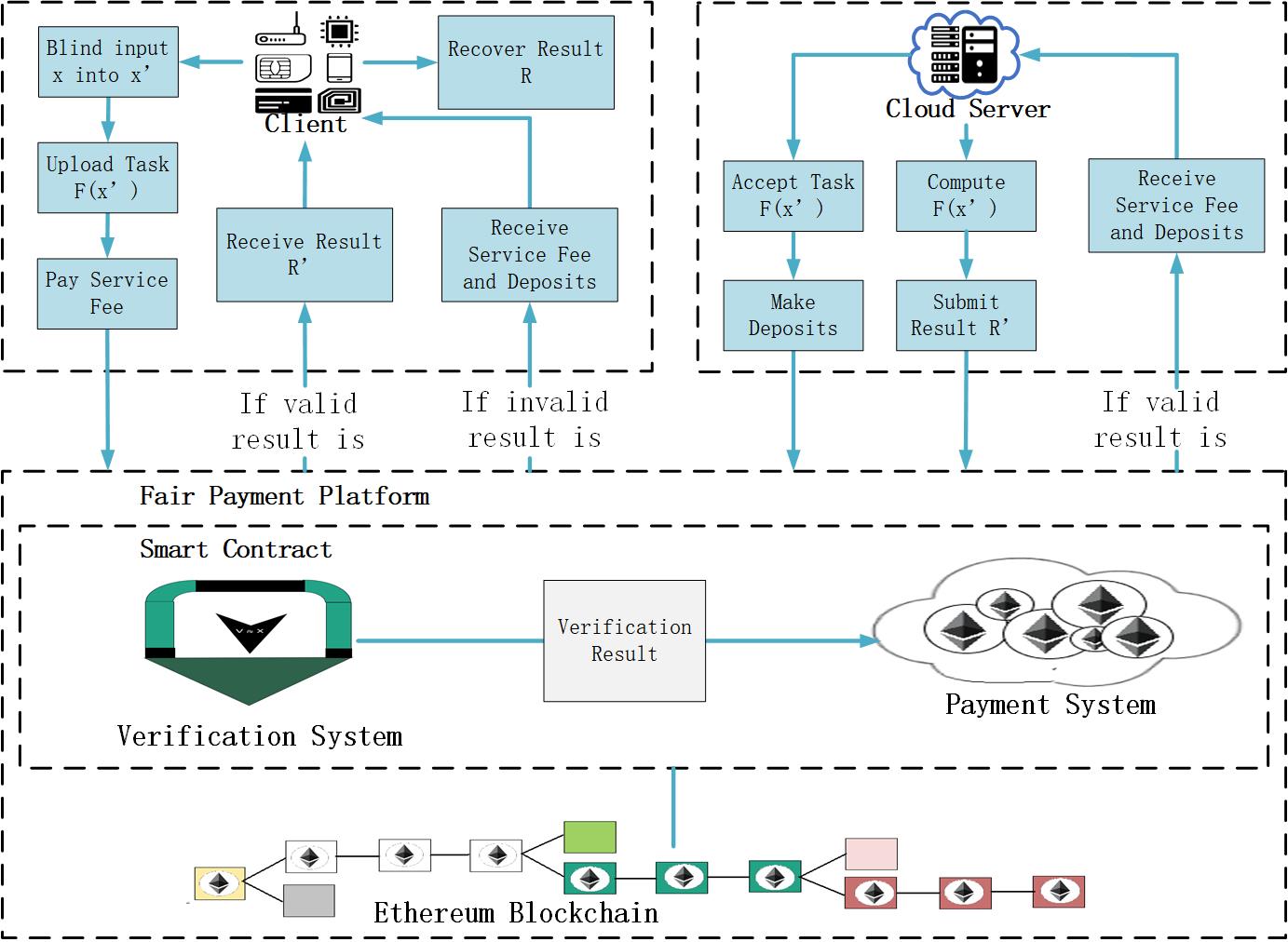}
	\caption{System Model}
	\label{fig:system model}
\end{figure*}

\subsection{Linear Regression}
Linear Regression is a regression model that can make regression forecasting in machine learning. It predicts a numerical value as accurately as possible through learning a linear model. There are variety of applications such as predicting the fuel efficiency of a car in terms of cylinders, displacement, horsepower, weight, etc..
Given a training set $D = \{(x_1,y_1), (x_2,y_2), ..., (x_n,y_n)\}$, where $x_i\in \mathbb{R}^{1\times m},$ $y_i\in\mathbb{R}$.
A typical linear regression model in machine learning is defined as:
\begin{align}
   y=X\omega,
\end{align}
where $y\in D$ is the vector $(y_1,y_2,...,y_n)^T$, $\omega\in \mathbb{R}^{n\times 1}$ is the vector $(\omega_1,\omega_2,...,\omega_n)^T$, $X\in D$ is a $m\times n$ matrix $[x_1,x_2,...,x_m]^T$ which is the $i$th row $x_i=(x_{i1},x_{i2},...,x_{in})$.
In this model, each vector $x_i$ in $X$ is taken as an input and the scalar $y_i$ is an output corresponding to $x_i$. The $\omega_i$ is the coefficient of $x_{ij}(1\le j\le n)$. We view it as a set of weights that determines the degree of prediction accuracy. The optimal coefficient $\omega$ is calculated as:
\begin{align}
	\omega&=(X^{T}X)^{-1}X^{T}y
\end{align}
In eq.(2), We need to execute a matrix inversion and two matrix multiplications. In the age of big data, the scale of training set generally becomes increasingly enormous. The client is not able to conduct such expensive computation locally especially for the resource-constraint client.

\subsection{Blockchain and Ethereum Smart Contract}
Since Nakamoto $et$ $al.$ first presented Bitcoin system in 2008~\cite{nakamoto2008bitcoin}, blockchain has attracted extensive attention of many researchers and enterprises. The blockchain is a decentralized shared ledger which is composed of many blocks in terms of chronological order and uses cryptography to guarantee tamper-resistance, traceability and unforgetablity. In Bitcoin, each block is composed of the block header and the block body. The block header describes the information of the block including Block Version, Time Stamp, Nonce, Parent Block Hash, Difficult and Merkle Tree Root Hash. The block body is a set of transactions in the block.

Nick $et$ $al.$ first proposed the conception of smart contract in 1995~\cite{szabo1997formalizing}. Smart contract is a digital protocol that is aimed at propagating, verifying or executing in an information way. Ethereum is the first platform that permits the developers to deploy their own smart contract~\cite{wood2014ethereum}. It provides a smart contract programming language named Solidity and a smart contract execution environment named Ethereum Virtual Machine (EVM). EVM is the core innovation of Ethereum and is a Turing complete software that runs on the Ethereum network. Smart contract developers write smart contract codes in Solidity and deploy it on Ethereum blockchain. Then the smart contracts are saved in a block. The smart contracts will be executed automatically only if the smart contract receives a specific trigger condition. Smart contract execution results are verified by all the Ethereum nodes and are stored on Ethereum blockchain.

\section{System Model and Definitions} \label{sec:three}
In this section, we first provide the system model. Then, we describe the security definitions.
\subsection{System Model}
Fig.~\ref{fig:system model} shows the system model. As we can observe, a secure and fair outsourcing scheme includes three entities: the client $C$, the cloud server $CS$ and the fair payment platform $FPP$. $C$ cannot carry out the heavy computation tasks with computation-constraint devices. Thus, he/she outsources computation tasks to $CS$. It is an untrustworthy entity that offers computation services. $FPP$ ensures the fairness of transaction between $C$ and $CS$. $FPP$ has two subsystems including a verification system and a payment system. The verification system is used to verify results from $CS$ while the payment system is used to ensure the fairness of payment.

The workflow of the proposed scheme is defined the following process: $C$ needs to leave computation tasks $F(x)$ to $CS$. First, $C$ blinds the input $x$ into $x^'$ and uploads the computation task $F(x^')$ to $FPP$. Meanwhile, $C$ pays the service fee to $FPP$. $CS$ accepts $F(x^')$ on $FPP$ and makes deposits to $FPP$. Then, $CS$ computes $F(x^')$ and submits the result $R^'$ to $FPP$. $FPP$ verifies $R^'$ from $CS$. If $R^'$ is an invalid result, $FPP$ transfers the service fee and the deposits to $C$. Otherwise, $FPP$ transfers the service fee and the deposits to $CS$ and sends $R^'$ to $C$. After receiving $R^'$ from $FPP$, $C$ recovers the real result $R$ from $R^'$.
\subsection{Security Definitions}
We introduce some security definitions for secure outsourcing computation including  framework, privacy, checkability and efficiency. Researchers have similar definitions on security properties of secure outsourcing computations \cite{gennaro2010non,10.1007/978-3-540-30576-7_15}, in which security, verifiability and efficiency are included. According to their theories, we summarize that a secure outsourcing algorithm satisfies the following properties:\\
\indent \textbf{Definition 1:}  A secure outsourcing computation algorithm $SOC$ = (\textit{\textbf{KeyGen, ProbGen, Compute, Verify, Recover}})
contains five algorithms defined belows.
\begin{itemize}
	\item \textit{\textbf{KeyGen}}($F$, $\lambda$) $\to$ ($PK$, $SK$): Given the security parameter $\lambda$, the randomized \textit{\textbf{KeyGen}} algorithm generates a public key $PK$ used to encode the target function $F$ and a secret key $SK$, which is used to obfuscate to computation inputs.
	\item $\textit{\textbf{ProbGen}}_{SK}(x) \to (\sigma_x, \tau_x)$: Using the secret key $SK$, the \textit{\textbf{ProbGen}} algorithm encodes the function input $x$ as a public value $\sigma_x$ which is submitted to server, and a secret value $\tau_x$ which is kept private by the
	client.
	\item $\textit{\textbf{Compute}}_{PK}(\sigma_x) \to \sigma_y$: Using the public key $PK$ and the encoded value $\sigma_x$, the server computes an encoded version $\sigma_y$ of the function’s output y = $F(x)$.
	\item $\textit{\textbf{Verify}}_{SK}(\tau_x, \sigma_y) \to 1 \bigcup 0$: Given the secret key $SK$ and the secret decoding $\tau_x$, the \textit{\textbf{Verify}} algorithm checks the correctness of $\sigma_y$. If the encoded output $\sigma_y$ is valid, this algorithm outputs 1. Otherwise it outputs 0.
	\item $\textit{\textbf{Recover}}(\tau_x, \sigma_y) \to y$: Using the secret key $SK$, the secret value $\tau_x$ and the encoded answer $\sigma_y$, the \textit{\textbf{Recover}} algorithm recovers the original result y = $F(x)$.
\end{itemize}
\indent \textbf{Definition 2} (Privacy \cite{gennaro2010non}): Privacy requires server cannot get any sensitive information in terms of the encoded input/output from client. We consider the following experiment:
\begin{align*}
(PK,SK)&\stackrel{R}{\gets}KeyGen(F,\lambda); \\
(x_0,x_1)&\gets \mathcal{A}^{PubProGen_{SK}(\cdot)}(PK);\\
(\sigma_0,\tau_0)&\gets ProGen_{SK}(x_0);\\
(\sigma_1,\tau_1)&\gets ProGen_{SK}(x_1);\\
b&\gets\{0,1\};\\
b^{'}&\gets \mathcal{A}^{PubProGen_{SK}(\cdot)}(PK,x_0,x_1,\sigma_b)
\end{align*}
\indent In the experiment, the adversary $\mathcal{A}$ is able to request the Oracle on any input he desires. The oracle $PubProbGen_{SK}(x)$ executes $ProbGen_{SK}(x)$ to generate $(\sigma_x,\tau_x)$ and returns only the public part $\sigma_x$.\\

For a secure outsourcing computation algorithm $SOC$, it is defined that advantage of an adversary $\mathcal{A}$ in the experiment as below:
\begin{equation*}
Adv^{C^S}_{\mathcal{A}}(F,\lambda) = \arrowvert Prob[b = b^'] - \frac{1}{2} \arrowvert
\end{equation*}
We define that a secure outsourcing computation algorithm $SOC$ is privacy if for any probabilistic polynomial time adversary $\mathcal{A}$,
\begin{equation*}
Adv^{C^S}_{\mathcal{A}}(F,\lambda) \leqslant negli(\lambda),
\end{equation*}
where negli() is a negligible function of its input.\\
\indent \textbf{Definition 3} ($\alpha$--Efficient \cite{10.1007/978-3-540-30576-7_15}): A pair of algorithms $(C,CS)$ is considered to be an $\alpha$-efficient execution of an algorithm $A$ if (1) the client and the cloud server correctly execute the algorithms and (2) for any inputs $x$, the execution time of $C^{CS}$ is less than or equal to an $\alpha$-multiplicative factor of the execution time of $A(x)$.\\
\indent \textbf{Definition 4} ($\beta$-checkable \cite{10.1007/978-3-540-30576-7_15}): A pair of algorithms $(C,CS)$ is considered to be a $\beta$-checkable execution of an algorithm $A$ if (1) the client and the cloud server correctly execute the algorithms and (2) for any inputs $x$ if a vicious server $CS^'$ depart from its preinstall functionality during the execution of $C^{CS^{'}}(x)$, $C$ will catch the error with probability greater than or equal to $\beta$.

\section{Proposed Scheme} \label{sec:four}
Firstly, we express the design rationale. Then, we describe our proposed scheme in detail.
\subsection{Design Rationale}
Our idea is to devise a novel scheme which allows client to securely perform linear regression model on cloud server. According to eq.(2), matrix inversion and matrix multiplications are the most time-consuming operations. Thus, we consider to outsource $(X^TX)^{-1}X^T$, and leave the matrix-vector production to be calculated locally. To maintain the confidence of inputs, we consider to apply a series of elementary transformations to $X$. By doing so, the position and the value of each element in $X$ and $X^T$ will be fully obscured. To guarantee the fairness of the outsourcing scheme, we develop a fair payment platform based on the blockchain. Client uploads computation tasks to the platform and cloud server accepts the computation task platform from the platform. The blockchain verifies the outputs which calculated by cloud server and guarantees the fairness. However, blockchain is a public ledger that anyone can have access to the data on it. Thus, we design a verification mechanism which does not involve any private information of the client. In other words, the proposed outsourcing scheme is publicly verifiable.

\subsection{Detailed Scheme}

 \begin{algorithm}
	\caption{The EFP-SOLR algorithm}
	\begin{algorithmic}
		\Require
		$X\in \mathbb{R}^{m\times n}$, a large-scale matrix; $X^T\in \mathbb{R}^{n\times m}$, the transpose of $X$; $y\in \mathbb{R}^{m\times 1}$, a vector.
		\Ensure
		 $\omega \in \mathbb{R}^{n\times 1}$, a vector such that $\omega=Ry$.	
		\State\textbf{Step 1.} \textbf{\textit{KeyGen}}($\lambda$)$\to$($SK$):
		\begin{itemize}
			\item Client chooses 2$k$ elementary transformation matrices:  $P_1,P_2...P_k,Q_1,Q_2,...Q_k$.
			\item Client sets two secret keys as follows:
				\begin{align}	
				SK_P&=\{{P_1,P_2,...,P_k}\}\label{gs:skp} \\
				SK_Q&=\{{Q_1,Q_2,...,Q_k}\}	\label{gs:skq}
				\end{align}
		\end{itemize}
	 	\State\textbf{Step 2.} $\textit{\textbf{ProbGen}}_{SK}(X,X^T) \to (X_1,X_2)$:
	 	\begin{itemize}
	 		\item Client computes $X_1$ and $X_2$ as:
	 		\begin{align}
	 		X_1&=XP_1P_2...P_k \label{gs:X1}\\
	 		X_2&=Q_{k}...Q_2Q_1X^{T}\label{gs:X2}
	 		\end{align}
	 	\end{itemize}
	 	\State\textbf{Step 3.} $\textit{\textbf{Compute}}(X_1,X_2) \to (R^')$:
	 	\begin{itemize}
	 		\item Cloud server calculates $R^'$ as follows:
	 		\begin{align}
	 		R^{'}=(X_2X_1)^{-1}X_2 \label{gs:R'}
	 		\end{align}
	 	\end{itemize}
	 	\State\textbf{Step 4.} $\textit{\textbf{Verify}}(R^') \to 1 \bigcup 0$:
	 	\begin{itemize}
	 		\item Fair payment platform chooses a random vector $r \in \mathbb{R}^{1 \times n}$ and calculates $V_1,V_2$ as:
	 		\begin{align}
	 		V_1&=rX_2\\
	 		V_2&=V_1X_1
	 		\end{align}
	 		\item Fair payment platform checks whether the following equation holds to verify validity of the result.
	 		\begin{align}
	 		V_1=V_2R^{'} \label{gs:V1_V2}
	 		\end{align}
	 	\end{itemize}
	 	\State\textbf{Step 5.} $\textit{\textbf{Recover}}(SK,R^',y) \to \omega$:
	 	\begin{itemize}
	 		\item Client calculates $R$ with the secret key $SK_P$ as:
	 		\begin{align}
	 		R=P_1P_2...P_kR^{'}
	 		\end{align}
	 		\item Client calculates the real output $\omega$ with $R$ as:
	 		\begin{align}
	 		\omega=Ry
	 		\end{align}
	 	\end{itemize}
	\end{algorithmic}
\end{algorithm}

Algorithm 1 shows our detailed proposed scheme $EFP-SOLR$. To protect the privacy of $X$ and $X^T$, client uses 2$k$($3<k \ll min(m,n)$) $n\times n$ elementary transformation matrices to blind $X$ and $X^T$. These elementary transformation matrices conduct the following three types of operations to a matrix:
\begin{itemize}
	\item \textbf{$Multiplication$}: A multiplication operation multiplies the $i$-th row (resp. column) of a matrix by a non-zero scalar.
	\item \textbf{$Permutation$}: A permutation operation makes the two rows (resp. columns) of a matrix exchange their location.
	\item \textbf{$Addition$}: An addition operation makes a matrix add $j$-th row (resp. column) multiplied by a non-zero scalar to the $i$-th row (resp. column).
\end{itemize}

\indent These elementary transformation matrices are invertible and their inverse matrices are easy to be calculated. In our scheme, $P_1$ and $Q_1$ are the elementary matrix of $Multiplication$ operations. Client chooses 2$n$ random scalars $p_1, p_2, ..., p_n, q_1, q_2, ..., q_n$ and constructs the following elementary transformation matrices:
  \begin{equation}
  	P_1=\begin{bmatrix}
  			p_1\\
  			&p_2 & &\\
  			&& \ddots\\
  			&&& p_n
  		\end{bmatrix},
  	Q_1=\begin{bmatrix}
  			q_1\\
  			&q_2 & &\\
  			&& \ddots\\
  			&&& q_n
  		\end{bmatrix}
  \end{equation}
$P_2$ and $Q_2$ are the elementary matrix of $Permutation$ operations. Client randomly generates a permutation $\pi_1$ and constructs $P_2$ as:
\begin{itemize}
		\item For each row $i$ in $P_2$, the value of $\pi_1(i)$-th element is 1 and the value of the other elements is 0.
\end{itemize}
We use a 4$\times$4 matrix to show the construction process of $P_2$. We assume permutation $\pi_1$ is as:
\begin{equation}
\pi_1=\begin{pmatrix}
1&2&3&4\\
3&4&2&1\\
\end{pmatrix}
\end{equation}
\begin{table*}[ht]
	\centering
	\fontsize{6.5}{8}\selectfont
	\begin{threeparttable}
		\caption{Time cost of phases.}
		\begin{tabular}{ccccccc}
			&\textit{ProbGen}&\textit{Compute}&\textit{Verify}&\textit{Recover}\cr
			\midrule
			SM&$2(k-1)mn+(k-2)(m+n)$& $m^2n+n^3+mn^2$&$3mn$&$kmn+(k-2)n$&\cr
			AS&$2mn$& $0$&$0$&$mn$&\cr
			\bottomrule
		\end{tabular}
	\end{threeparttable}
	\label{table:timecost}
\end{table*}
Then, $P_2$ is constructed as:
\begin{equation}
P_2=\begin{bmatrix}
0&0&1&0\\
0&0&0&1\\
0&1&0&0\\
1&0&0&0\\
\end{bmatrix}
\end{equation}
The construction process of $Q_2$ is similar to $P_2$. $P_3,...,P_k,Q_3,...,Q_k$ are the elementary matrices of $Addition$ operations. Client chooses $2k-4$ random scalars $r_1,r_2,...,r_{2k-4}$ and constructs $P_3$ as:
\begin{itemize}
	\item Set the value of a randomly chosen element which is not on the main diagonal as $r_1$.
	\item Set the value of each element on the main diagonal as  1.
	\item Set the values of other elements as 0.
\end{itemize}
For example, we use a 4$\times$4 matrix to show the construction process of $P_3$.  Assuming $r_1=3$, constructe $P_3$ as follows:
\begin{equation}
P_3=\begin{bmatrix}
1&0&3&0\\
0&1&0&0\\
0&0&1&0\\
0&0&0&1\\
\end{bmatrix}
\end{equation}
\indent The constructions of $P_4,...,P_k,Q_4,...,Q_k$ share the same logic with $P_3$. Then the client preserves these elementary transformation matrices as privacy keys. To blind the computation input $X$ and $X^T$, the client computes $X_1$ and $X_2$ as in eq.(5) and eq.(6). Note that the eq.(5) is calculated from left to right, while the eq.(6) is calculated from right to left. Then, client uploads the computation task $F(X_1,X_2)$ and pays service fee to fair payment platform. Cloud server accepts $F(X_1,X_2)$ on fair payment platform and makes deposits to fair payment platform. Then, the cloud server performs the computation task as in eq.(7). After solving $F(X_1,X_2)$, the cloud server submits the result $R^'$ to the fair payment platform. On receiving $R^'$ from cloud server, fair payment platform chooses a random vector $r \in \mathbb{R}^{1 \times n}$ and calculates $V_1$, $V_2$ as in eq.(8) and  eq.(9). Then the fair payment platform inspects the validity of $R^'$ by checking whether the eq.(10) holds. If the result $R^{'}$ is valid, fair payment platform transfers the service fee and the deposits to cloud server and stores the result $R^'$. Otherwise, fair payment platform transfers the service fee and the deposits to the client. The client downloads the valid result $R^'$ on fair payment platform and recovers the result $\omega$ as $\omega = P_1P_2...P_kR^{'}y$.

\begin{remark}
Note that we use the elementary transformation matrices to illustrate the elementary transformations. In fact, when implementing the scheme, we directly conduct elementary transform a matrix instead of multiplying the matrix by a elementary transformation matrix. The reason is that multiplying the input matrix by a elementary transformation matrix will produce many unnecessary extra scalar multiplications. For example, if we multiply the input $X$ by $P_3$ in eq.(15), the elements of $X$ will be multiplied by the elements 0 and 1 in $P_3$. However, these scalar multiplications are unnecessary and time-consuming. Thus, we directly add the elements of the first column multiplied by three to the elements of the third column.
\end{remark}

\subsection{Correctness Analysis}
We denote $P$ as $P = P_1P_2...P_k$ and $Q$ as $Q = Q_k...Q_2Q_1$, respectively. Notice that the client recovers the results $R$ from $R'$ as $R = P_1P_2...P_kR'$, where
\begin{align*}
P_1P_2...P_kR' &= PR^{'}\\
&=P(X_2X_1)^{-1}X_2\\
&=P(QX^{T}XP)^{-1}QX^{T} \\
&=PP^{-1}(X^{T}X)^{-1}Q^{-1}QX^{T} \\
&= (X^{T}X)^{-1}X^{T}
\end{align*}
Thus, the recovered $R$ is the original computation output. We now prove the correctness of the verification process. Blockchain inspects the validity of $R'$ by checking whether $V_1=V_2R^'$, where
\begin{align*}
V_2R^{'}&=rX_2X_1(X_2X_1)^{-1}X_2 \\
&=rX_2 \\
&=V_1
\end{align*}
Thus, if $V_1=V_2R^{'}$, the returned result is correct.

\subsection{Security Analysis}

\indent \textbf{Theorem 1:} \textit {The presented algorithms $EFP-SOLR$ is input/output private.}\\
\indent \textbf{Proof.} We first analyze the privacy of inputs in our scheme, which are $X$ and $X^T$. In our scheme, the client transforms $X$ into $X_1$ via multiplying a series of elementary transformation matrices. Note that these elementary transformation matrices are randomly chosen. Each element in $X$ is obscured completely. Thus, cloud server is not able to retrieve $X$ from $X_1$ without the series of elementary transformation matrices. Notice that $X^{T}$ is also obscured by the elementary transformation matrices. We assume that the cloud server can correctly guess an element of input/output matrices with a probability of $\frac{1}{\delta}$. In fact, the probability is very small because the elements of input/output matrices are real number. Therefore, for the inputs $X$ and $X^T$, the probability that cloud could correctly guess the input $X$ or $X^T$ is $negli(mn)=\frac{1}{\delta^{mn}}$. For the output $R$, $R$ is equal to $PR^'$. The cloud server can reveal the real result $R$ only if the cloud server gets the private key $SK_P$. However, the $SK_P$ is stored at the client side. The probability that cloud is able to correctly guess the result $R$ is $negli(mn)=\frac{1}{\delta^{mn}}$. As we all have observed, for any valid input ($X,X^T$) and output ($R$), the probability that cloud could correctly guess the inputs or outputs both are nonpolynomial-time function. Thus, the probability that the cloud server reveals input ($X,X^T$) or output ($R^'$) is negligible.

\subsection{Efficiency Analysis}
\indent \textbf{Theorem 2:} \textit {The proposed algorithm $EFP-SOLR$ is $(3(k-1)mn+(k-2)(m+2n))/(m^2n+n^3+mn^2)$ efficient.}\\
\indent \textbf{Proof.} We analyze the efficiency of our scheme as shown in Table I. We assume scalar multiplication as $SM$ and assignment operation as $AS$. We ignore calculating light computations such as scalar addition. Actually, $AS$ consumes less time than $SM$. In the step \textit{ProbGen}, client executes $2(k-1)mn+(k-2)(m+n)$ $SM$ and $2mn$ $AS$. In the step \textit{Compute}, cloud server executes $m^2n+n^3+mn^2$ $SM$. In the step \textit{Verify}, fair payment platform executes $3mn$ $SM$. In the step \textit{Recover}, client executes $(k-1)mn+(k-2)n$ $SM$ and $mn$ $AS$. Therefore, client executes total $3(k-1)mn+(k-2)(m+2n)$ $SM$ and $3mn$ $AS$. Cloud server executes total $m^2n+n^3+mn^2$ $SM$. The complexity of client executing without outsourcing algrithm is as high as the complexity of cloud server executing the step \textit{Compute}. Fair payment platform executes $3mn$ $SM$. Thus, according to definition 3, $EFP-SOLR$ is ($(3(k-1)mn+(k-2)(m+2n))/(m^2n+n^3+mn^2)$) efficient.

\indent \textbf{Theorem 3:} \textit{ The proposed algorithms $EFP-SOLR$ is 100 \%-verifiable secure outsourcing algorithm}\\
\indent \textbf{Proof.} According to definition 4, we need to prove that
if a cloud server performs malicious behavior, the probability that fair payment platform can detect the malicious behavior is 1. Fair payment platform performing $Verification$ sub-algorithm needs four parameters including $r$, $X_1$, $X_2$ and $R^'$. Parameter $r$ is produced by himself. Parameter $X_1$ and $X_2$ are produced by client. Only parameter $R^'$ is produced by cloud server. Thus, fair payment platform can completely detect the malicious behavior by checking if the eq.(10) not holds.
\indent \textbf{Communication cost.} Table. \ref{table:Comcost}. shows communication cost of $EFP-SOLR$ alogorithms. In $EFP-SOLR$ algorithm, client sends the blind matrix $X_1\in \mathbb{R}^{m\times n}$ and $X_2\in \mathbb{R}^{n\times m}$ to cloud server in the phase $ProbGen$. In the phase $Compute$, cloud server computes the result $R^{'}\in \mathbb{R}^{n\times m}$ in terms of $R^{'}=(X_2X_1)^{-1}X_2$. And then, cloud server returns $R^'$ to client. Client conducts phase $Verify$ and $Recover$ locally. Thus, there is no communication cost in $Verify$ and $Recover$.
\begin{table}[h]
	\centering
	\begin{threeparttable}
		\caption{Communication cost}
		\begin{tabular}{ccccp{10 pt}p{0.1 pt}p{0.1 pt}}
			&$EFP-SOLR$&\cr
			\midrule
			\textit{ProbGen}&$C\xrightarrow{X_{1},X_{2}} CS$&\cr
			\textit{Compute}& $CS\xrightarrow{R^{'}} C$&\cr
			\textit{Verify}&0&\cr
			\textit{Recover}&0&\cr
			\bottomrule
		\end{tabular}
		\label{table:Comcost}
	\end{threeparttable}
\end{table}

\section{System Implementation} \label{sec:six}
In this section, we introduce the implementation details of the developed system. We first describe the system architecture, then we show an example of the developed smart contract.
\begin{figure}[htb]
	\includegraphics[width=1\linewidth]{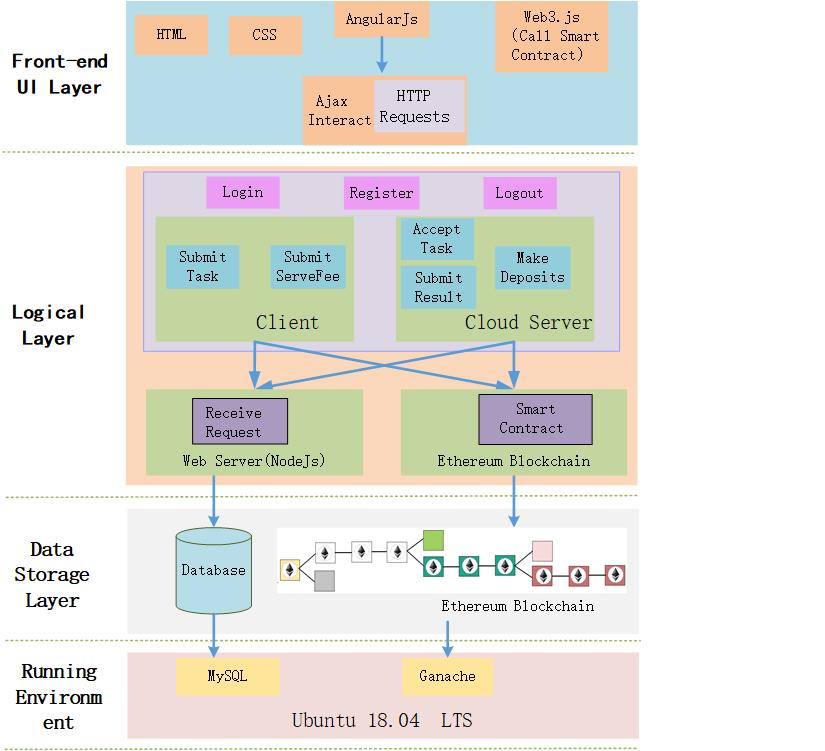}
	\caption{System Implementation Architecture}
	\label{fig:system architecture}
\end{figure}

\subsection{System Implementation Architecture Diagram}
Fig.~\ref{fig:system architecture} shows the architecture of the developed fair payment system, which consists four layers. The front-end UI layer provides the graphical user interface for users. We develop a series of scripts written in AngularJS to realize ajax interaction. Ajax uses HTTP requests to realize asynchronous data transmission between the browser and the web server. Web3.JS is a JavaScript library provided by Ethereum, which provides the interface for front-end javascripts to interact with the smart contract functions. We use MetaMask to manage the ethereum account. MetaMask is a plug-in wallet for web browsers that allows users to interact with the Ethereum blockchain; The logical layer includes a series of functions. It receives requests from UI layer and executes the corresponding functions. These functions are provided by the web server and smart contract. We develop the web server in NodeJS and develop the smart contract in Solidity. The web server interacts with the database while the smart contract interacts with the Ethereum Blockchain; Data Storage Layer provides data interaction to the Logical layer. The data is stored in database or Ethereum Blockchain. Specifically, it stores the basic information (e.g. user-name, user-password and user-ID) in the database, and stores crucial calculation related data (e.g. task-parameters and server fee) on the blockchain. Our fair payment system runs on Ubuntu 18.04 LTS. We use Ganache to simulate Ethereum Blockchain, which is a private Ethereum blockchain for developers that can be used for local deployment of smart contracts.

\begin{figure*}
 \subfigcapskip=5pt
 \centering
 \subfigure[Submit Task]{
 \begin{minipage}{0.2\linewidth}
  \centering
  \includegraphics[height=1.8in,  width=1.2in]{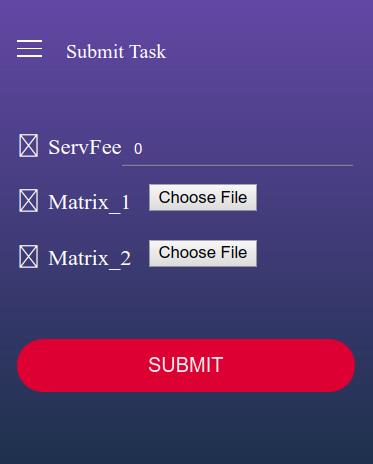}
 \end{minipage}%
 }%
 \subfigure[Transaction]{
 \begin{minipage}{0.2\linewidth}
  \centering
  \includegraphics[height=1.8in,width=1.2in]{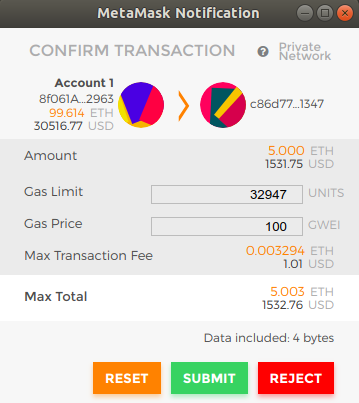}
 \end{minipage}%
 }%
 \subfigure[Tasks List]{
 \begin{minipage}{0.2\linewidth}
  \centering
  \includegraphics[height=1.8in,width=1.2in]{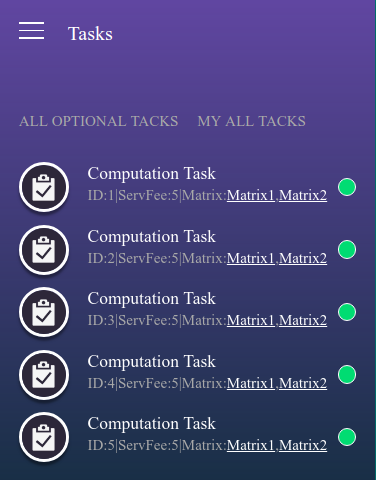}
 \end{minipage}%
 }%
 \subfigure[Claim Task]{
 \begin{minipage}{0.2\linewidth}
  \centering
  \includegraphics[height=1.8in,width=1.2in]{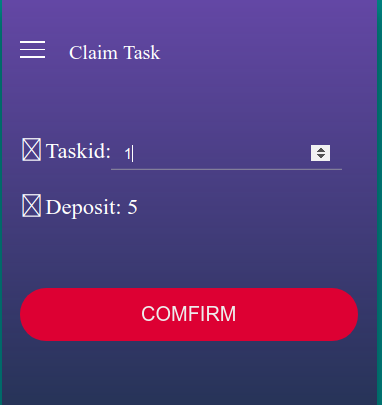}
 \end{minipage}%
 }%
 \subfigure[Submit Result]{
 \begin{minipage}{0.2\linewidth}
  \centering
  \includegraphics[height=1.8in,width=1.2in]{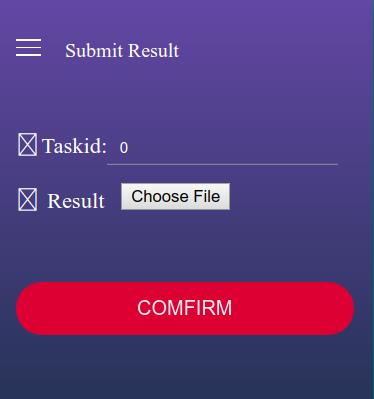}
 \end{minipage}%
 }%
 \centering
 \caption{System Demonstration}
 \label{fig:system yanshi}
\end{figure*}
\begin{lstlisting}[language=C,title=Listing. 1: Solidity code of the Payment function]
function Payment(uint taskid,
uint CID, uint8 flag) payable{
  var task = Tasks[taskid];
  address addr;
  uint e = 1 ether;
  uint SerPlusDep = task.ServFee +
  task.Deposit;
  if(flag == 1){
    task.status = 2;
    addr = task.Cloud_addr;
    addr.transfer(SerPlusDep*e);
  }else{
    task.status = 3;
    addr = task.Client_addr;
    addr.transfer(SerPlusDep*e);
  }
}
\end{lstlisting}
Listing. 1 shows the Solidity code of function $Payment$, which takes three input parameters including $taskid$, $CID$, $flag$. $taskid$ is the task ID and $CID$ is the cloud server ID. Flag $flag$ is the verification outcome, which determines whether the calculation result passed the verification. When the value of $flag$ is 1, smart contract will transfer the service fee and deposits to cloud server. When the value of $flag$ is 0, smart contract will transfer the service fee and deposits to client. The variable $task$ is a structural body which contains all the information of a task. $status$ is an attribute of $task$, which represents the status of $task$. When $status$ equals 0, it means that the task is uploaded by the client and has not been taken by any cloud server yet. When the value of $status$ is 1, it represents that the task has been taken but not solved yet. When the value of $status$ is 2, it means that the task has been correctly conducted by a cloud server. When the value of $status$ is 3, it means that a cloud server returned an invalid result of the task. $Tasks$ is a mapping which maps the variable type $uint$ to the variable type $struct$. $addr$ is a variable of type $address$, which represents an Ethereum address to receive ether. $e$ is an ether unit. $SerPlusDep$ represents the sum of service fee and deposits.

\subsection{System Demonstration}
Fig.~\ref{fig:system yanshi} demonstrates the developed system. To outsource a computation task, the client submits the computation task as shown in Fig.~\ref{fig:system yanshi}(a). The client input the service fee and two matrices. In this example the client pays five ethers for a LR computation task. When the client clicks the Submit button, MetaMask will package the transaction as shown in Fig.~\ref{fig:system yanshi}(b). In Fig.~\ref{fig:system yanshi}(b), client uses an ethereum address (0x8fO61A...2963) to transfer five ethers as serve fee to smart contract address (0xc86d77...1347). Client and cloud server can view all tasks in the task list as shown in Fig.~\ref{fig:system yanshi}(c). The task list shows all task information including Taskid, Serve Fee and Matrix. Client and cloud server can view detailed matrix by clicking on the following link. In Fig.~\ref{fig:system yanshi}(d), the cloud server input the task ID to claim a computation task. Then the fair payment system will show the deposits, which is equal to the service fee. When the cloud server clicks the COMFIRM button, MetaMask also packages the transaction, which is similar to Fig.~\ref{fig:system yanshi}(b). In this transaction, the cloud server uses an ethereum address (0xb2b04a...240B) to transfer five ethers as deposits to smart contract address (0xc86d77...1347). After completing the task, cloud server submits the result as shown in Fig.~\ref{fig:system yanshi}(e). Cloud server input the task ID and uploads a txt file which stores the result. After cloud server clicks the CONFIRM button, MetaMask still packages the transaction which is similar to Fig.~\ref{fig:system yanshi}(b). Finally, smart contract checks the validity of result and transfers deposits and service fee to the corresponding address.

\section{Experimental Performance Evaluation} \label{sec:seven}

In this section, we evaluate the practical performance of our proposed scheme. We first describe the evaluation methodology, then show the evaluation results.
\subsection{Evaluation Methodology}
In our experiments, we conduct the simulation of all phases of our scheme on a Windows machine. Specifically, the testbed is with the Windows 10 on an i5-6500 at 3.20GHz with 8GB memory. We use python to implement our proposed algorithms. We execute the experiment 20 times and calculate the average execution time.

\begin{figure*}[htbp]
 \subfigure[Time cost on the client]{
  \begin{minipage}[t]{0.34\linewidth}
   \includegraphics[width=2.4in]{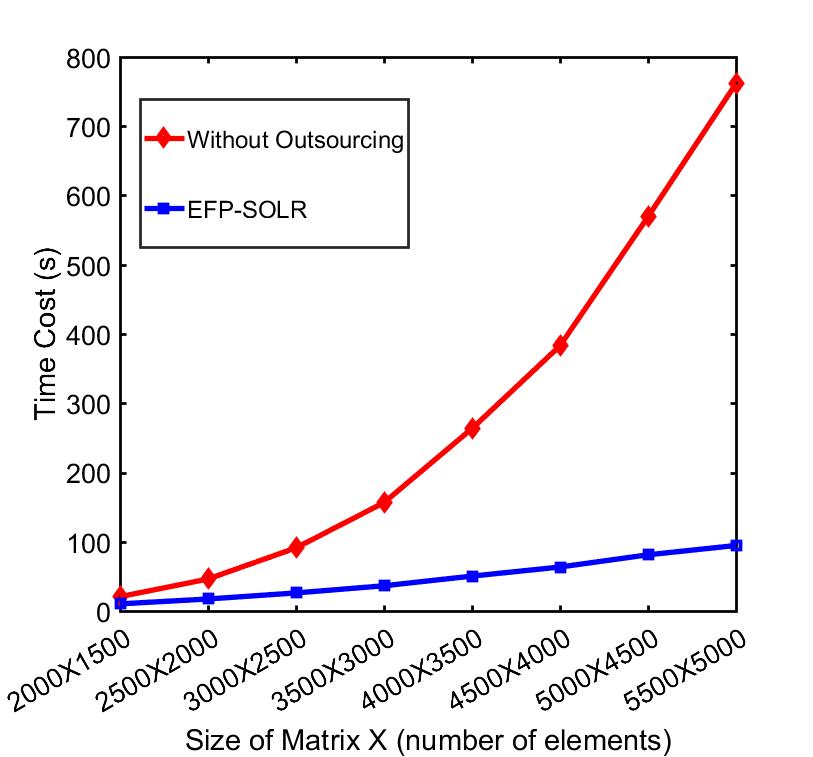}
  \end{minipage}%
 }%
 \subfigure[Time cost comparison among phases]{
  \begin{minipage}[t]{0.34\linewidth}
   \centering
   \includegraphics[width=2.4in]{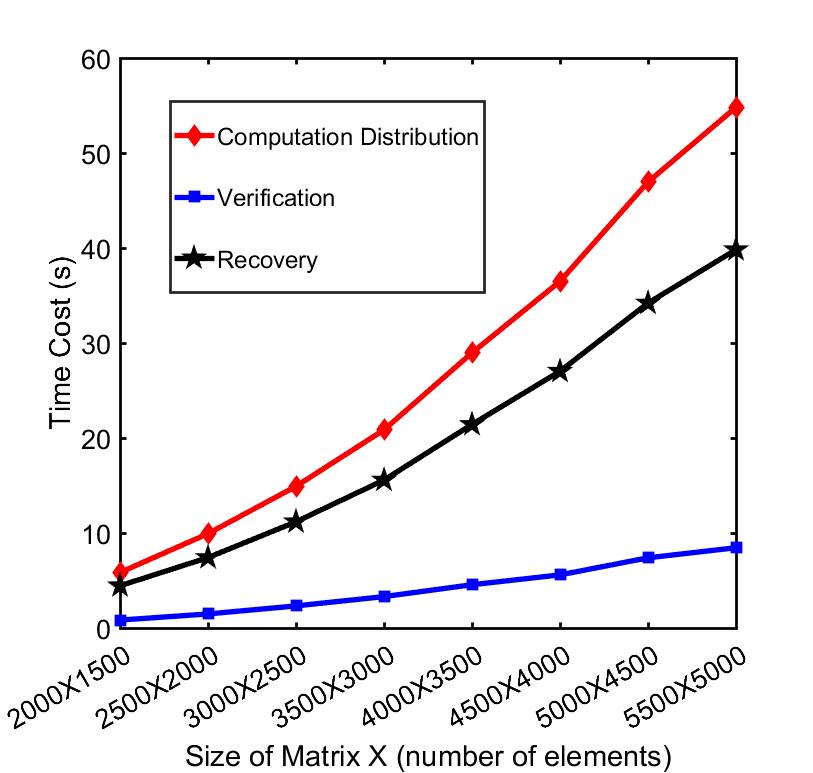}
  \end{minipage}%
 }%
 \subfigure[Time cost comparison among phases]{
  \begin{minipage}[t]{0.34\linewidth}
   \centering
   \includegraphics[width=2.4in]{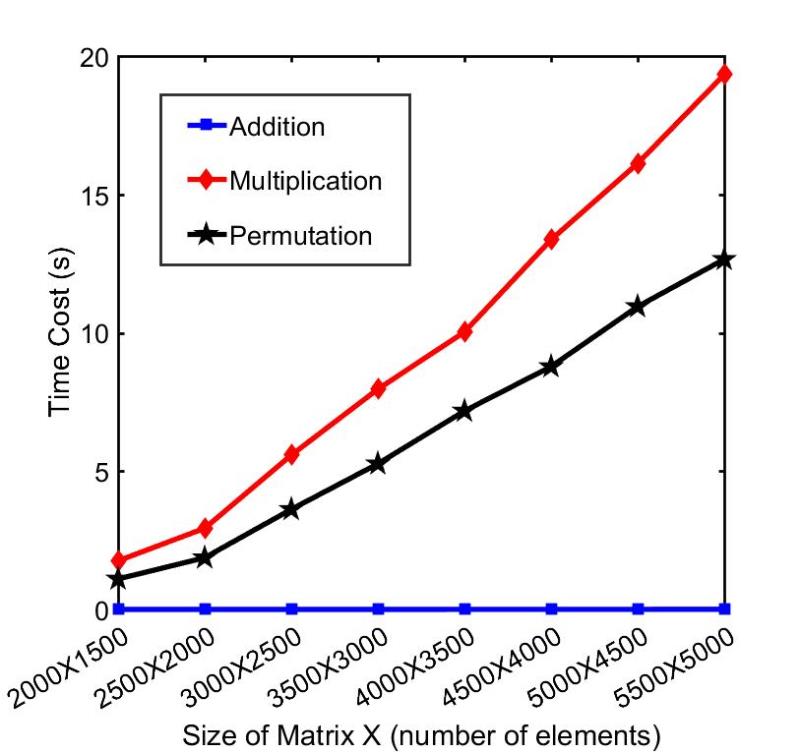}
  \end{minipage}%
 }%
 \caption{Evaluation results for $EFP-SOLR$}
 \label{fig:xiaolv}
\end{figure*}

\subsection{Evaluation Results}
We present the evaluation results for $EFP-SOLR$ in Fig.~\ref{fig:xiaolv}. The size of the matrix $X$ ranges from $2000 \times 1500$ to $5500 \times 5000$. Fig.~\ref{fig:xiaolv}(a) compares time cost between $EFP-SOLR$ and conducting the LR without-outsourcing at the client. We can observe that our proposed $EFP-SOLR$ costs much less time. Fig.~\ref{fig:xiaolv}(b) shows time cost comparison among different phases. Obviously phase $ProbGen$ is the most time-consuming. The reason is that as discussed in Section~\ref{sec:four}, the $ProbGen$ phase requires more scalar multiplications than other phases. In the Fig.~\ref{fig:xiaolv}(c), we compare the time cost among three elementary transformations. $Multiplication$ and $Addition$ are mainly composed of scalar multiplications, while $Permutation$ is mainly composed of assignment operations. $Multiplication$ performs much few scalar multiplications than $Addition$. Thus, there is a significant difference between $Multiplication$ and $Addition$. $Permutation$ consumes less time than $Multiplication$. The reason is that the assignment operation consumes less time than scalar multiplication.

\section{Applications}\label{sec:app}

In this section, we discuss possible applications where our proposed scheme can be applied. Linear regression has been applied in many applications, such as face recognition~\cite{Naseem2010Linear} and disease prediction~\cite{2020Prediction}. In these applications, data involved in machine learning often contain some sensitive information. Thus, when users use a cloud server to accomplish time-consuming machine learning, data security faces critical challenges. To protect the data security while leveraging the cloud server, our proposed scheme can be applied.

\begin{figure}[htb]
	\includegraphics[width=1\linewidth]{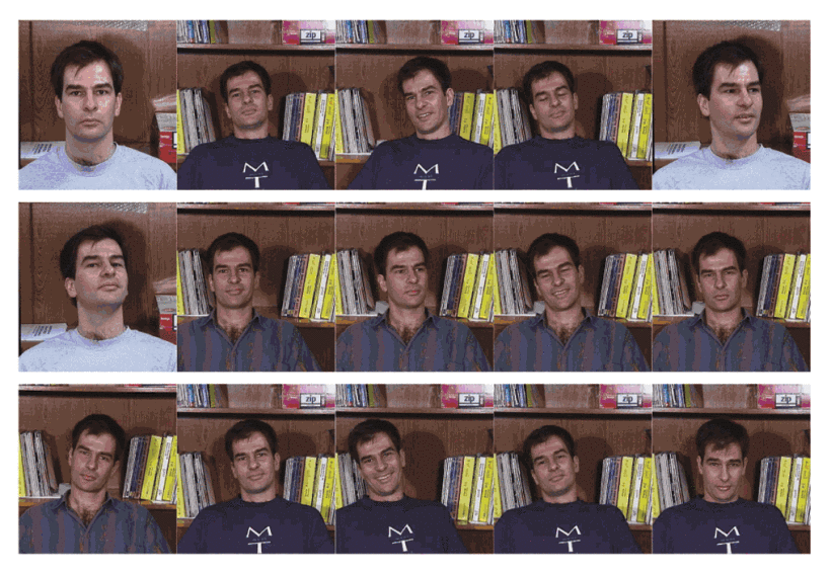}
	\caption{Georgia Tech face database \cite{Naseem2010Linear} }
	\label{fig:face}
\end{figure}

For example, Naseema {\em et al.} in~\cite{Naseem2010Linear} proposed a novel approach of face identification by formulating the pattern recognition problem in terms of linear regression. The data-set is shown is Fig.~\ref{fig:face}. The features of a person's face are supposed to be private information, especially when the face identification is applied in military uses, the leak of the training data may cause serious losses. Thus, to ensure data security, we can apply our proposed secure outsourcing algorithm to accomplish the face identification. By doing so, the user can efficiently accomplish the face identification with the help of a cloud server, while the data privacy, result verifiability and payment fairness are ensured.

\begin{figure}[htb]
	\includegraphics[width=1\linewidth]{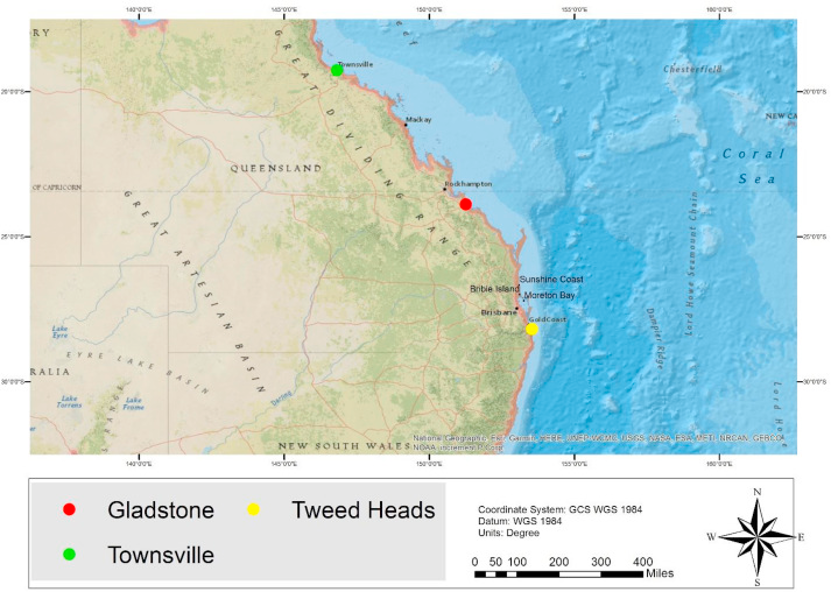}
	\caption{Map of Australia Coastal Sites \cite{0Near}}
	\label{fig:haian}
\end{figure}

Fig.~\ref{fig:haian} shows a map of an Australia coastal site. Ali {\em et al.} in \cite{0Near} applied linear regression to develop a real-time significant wave height forecasting system. The geography data involved in the forecasting system should be kept secret because of its commercial value. Thus, we can apply our proposed secure outsourcing algorithm to accomplish the significant wave height forecasting. Our algorithm can also be applied in many other applications where the linear regression is in used, so that the data security in those applications can be guaranteed.

\section{Related work} \label{sec:eight}
\subsection{Machine Learning And Linear Regression}
\indent Applying machine learning approaches to efficiently analyze large scale of matrix data has allured increased attention due to their special feature for facilitating pattern recognition, classification, and prediction. Regression and multilevel/hierarchical models are widely implemented to conduct data processing with linear or nonlinear regression and multilevel models \cite{hayes2013introduction, aiken1994multiple}. In the past, a number of machine learning and statistical methods have been proposed to generate meaningful information on different datasets~\cite{yamada2018asymptotic, zhang2019inductive, gao2018aiproannotator, fasoranbaku2018towards, tu2018multi, balasubramaniam2018understanding, luong2018a, zhul2018a, zhou2018the}. For example, Huang {\em et al.} \cite{huang2019improving} presented a novel approach to compute special label features of multiple label. They proposed a new augmented matrix using advanced order label correlations and implement a multi-label classifier simultaneously to enhance multi-label classification. Tu {\em et al.} \cite{tu2018multi} proposed a multiple label answer aggregation method which applied the Joint Matrix Factorization (JMF) to picky and mutually factorizes the sample label bond matrices collected from products of individual by different annotators. \\
\indent There are also many research efforts focusing on improving the efficiency and effectiveness of linear regression models \cite{goryaeva2019towards, lukmanand2016some, fasoranbaku2018towards, yamada2018asymptotic, fasoranbaku2018towards}. In a linear regression model, it is often assumed that the explanatory variables are independent. Lukmanand {\em et al.} \cite{lukmanand2016some} proposed estimators based on Hoerl and Kennard estimation techniques to improve the ridge parameter. Fasoranbaku {\em et al.} \cite{fasoranbaku2018towards} evaluated the basis of six parameters and helped to improve more powerful experiment appropriate for better parameters estimation of the LR. Chen {\em et al.} \cite{chen2014highly} presented an approach to securely performing linear regression on a cloud. However, their scheme can not hide the number of element 0 in the process of blinding inputs. Zhou {\em et al.} \cite{zhou2018efficiently} introduced a secure method to outsourcing linear regression. The proposed scheme protected the privacy of inputs and outputs.\\

\subsection{Secure Outsourcing Computations}
\indent There are extensive research efforts on a variety of secure outsourcing schemes for scientific computations. For example, Atallah {\em et al.} in \cite{APR02} first presented a generic structure for the secure outsourcing of scientific computations. However, their presented framework could not verify the correctness of the result which calculated by cloud server. Chen {\em et al.} in \cite{CHL15} designed a secure outsourcing approach for the large-scale linear equations. Their approach used some special sparse matrixes to blind the inputs and outputs. And the approach allowed client to detect cheating behavior of cloud servers with a probability of 100\%. Salinas {\em et al.} in \cite{SLC17} proposed a secure outsourcing method which allows resource-constrained devices to solve large-scale sparse linear systems of equations (SLSEs). The proposed method protected the privacy of inputs/outputs and is efficient comparing with other schemes simultaneously. The computation cost of some basic cryptographic operations is too heavy for resource-constrained devices. To free these devices from such computations, a number of research efforts have been conducted on how to securely outsource cryptographic computations \cite{CSL15,CP92,HL05,TZR15,CCM10}. Hohenberger {\em et al.} in \cite{HL05} proposed a security framework for outsourcing cryptographic computations. Based on the frameworkm, they proposed two practical outsource-secure approaches. Zhang {\em et al.} in \cite{zhang2020practical} proposed two practical algorithms to securely outsource the Cippola's algorithm. The proposed two schemes enable IoT devices to accomplish the Cippola's algorithm efficiently. Also, IoT devices can detect the misbehavior of  cloud servers with a probability of 1. Yu {\em et al.} in \cite{YRW16} designed a cloud storage auditing scheme which achieved the verifiable outsourcing of key updates. In the presented scheme, the key updates was be able to securely outsourced to an authorized entity, which reduced the key-update burden on the user.\\

\subsection{Blockchain}
\indent Blockchain technology enables secure, trusted, and decentralized autonomous ecosystems for various scenarios. The advanced blockchain technology has been widely leveraged in machine learning to securely and efficiently collect, organzine and audit the extensive quantities of data for model building and accurate prediction \cite{yuan2018blockchain, li2018block, wang2019blockchain, tian2019block, huang2017behavior, juneja2018leveraging, ide2018collaborative, kurtulmus2018trustless}. Li {\em et al.} \cite{li2018block} presented a security mechanics for distributed cloud storage based on blockchain. In their framework, client could distribute all their data into encrypted data blocks and send these data blocks randomly to blockchain network. Juneja {\em et al.} in \cite{juneja2018leveraging} implemented blockchain technology to develop an access control system in which classifier can safely store and access data during retraining in real-time using Stacked Denoising Autoencoders (SDA) networks. Kurtulmus {\em et al.} in \cite{kurtulmus2018trustless} proposed a blockchain-based model for exchanging machine learning models. They used the Ethereum blockchain to create contracts that offer a reward in exchange for a trained machine learning model for a particular data set.

Shafagh in {\em et al.} \cite{SLA17}  proposed a blockchain-based auditable storage and sharing scheme of IoT Data. Their proposed scheme provides distributed access control and data management. Different from existing trust model that delegates access control of our data to a centralized trusted authority, their proposed scheme empowers the users with data ownership. To provide a systematic review of blockchain technology in IoT, Christidis {\em et al.} \cite{christidis2016blockchains} explored how the combination of Blockchain and Internet of Thing (IoT). In their research, they stated that it will has a bright prospect and will lead to significant changes in multiple industries When Blockchain and IoT are combined. In this paper, we employ the blockchain as the middleman, which verifies the calculation results and guarantees the fairness and further ensure the security and accuracy of our outsourcing algorithm. Lin {\em et al.} in \cite{lin2020blockchain} proposed a blockchain-based system to securely outsource the billinear pairing. In the proposed system, the cloud server can get paid for the computation task only when he correctly performed the outsourced workload from the client.

\section{Conclusion and future work} \label{sec:nine}
In this paper, we designed a secure, verifiable and fair scheme to outsource a classical statistical machine learning models: the linear regression. Similar practices can be applied to other statistical machine learning models. The presented scheme prevents the computation input and output from leaking to cloud server, and the computation result is verifiable. Also, fairness is guaranteed by the blockchain. We introduced the presented approach detailedly and analyzed correctness, security and efficiency of it. In addition, we developed the fair, verifiable system on the Ethereum blockchain. To evaluate our presented scheme, we carried out some experiments. The experimental data indicate that our presented algorithm is efficient.

The cloud-aided machine learning faces security challenges, including data privacy, result verifiability and payment fairness. In this work, we studied the classic linear regression as an example to show how to address these challenges. To the best of our knowledge, there is no generic secure outsourcing approach for all machine learning algorithms. Fully homomorphic encryption (FHE) is a possible solution for a generic secure outsourcing approach, but the efficiency of the FHE is too low such that the FHE-based approach is not practical. Thus, current researches focus on designing specific outsourcing approaches for particular machine learning algorithms. In our future work, we plan to explore other specific machine learning algorithms, in which data privacy, result verifiability and payment fairness are guaranteed. To enable fair payment, we employ the blockchain technology, which plays a role in our scheme to verify the computation result and make a judgment accordingly. Notice that the blockchain is a public ledger that anyone can view the content. Thus, when designing such outsourcing algorithms, the task on the blockchain (the verification process) cannot involve any sensitive data. The outsourcing algorithms have to be publicly verifiable.

\section* {Acknowledgments}
This research is supported by National Natural Science Foundation of China (61572267), National Development Foundation of Cryptography (MMJJ20170118), Key Research and Development Project of Shandong Province (2019GGX101051).

\bibliographystyle{ieeetr}
\bibliography{ref}

\end{document}